\documentclass[10pt,a4paper,twocolumn]{article}
\usepackage[T1]{fontenc}
\usepackage[utf8]{inputenc}
\usepackage{lmodern}
\usepackage[margin=18mm]{geometry}
\usepackage{amsmath,amssymb}
\usepackage{graphicx}
\IfFileExists{booktabs.sty}{\usepackage{booktabs}}{\newcommand{\toprule}{\hline}\newcommand{\midrule}{\hline}\newcommand{\bottomrule}{\hline}}
\IfFileExists{microtype.sty}{\usepackage{microtype}}{}
\IfFileExists{hyperref.sty}{\usepackage[hidelinks]{hyperref}}{\newcommand{\href}[2]{#2}\newcommand{\nolinkurl}[1]{\texttt{#1}}\newcommand{\hypersetup}[1]{}}
\hypersetup{pdftitle={Strain-transfer at the Cu nanoparticle--Si(111) interface: static atomistic modelling and simulated Bragg CDI},pdfauthor={Corentin Chatelier}}
\newcommand{\angstrom}{\mbox{\AA}}
\newcommand{\doi}[1]{\href{https://doi.org/#1}{doi:\nolinkurl{#1}}}
\newcommand{\paperfigure}[1]{\IfFileExists{figures/#1.pdf}{\includegraphics[width=\linewidth]{figures/#1.pdf}}{\fbox{\parbox{0.91\linewidth}{\centering Figure: \texttt{\detokenize{#1}}}}}}
\title{Strain-transfer at the Cu nanoparticle--Si(111) interface: static atomistic modelling and simulated Bragg CDI}
\author{Corentin Chatelier\\\small Université Grenoble Alpes, CEA, Liten, DTNM, Grenoble 38000, France\\\small corentin.chatelier@cea.fr}
\date{14 October 2026}

\begin{document}
\twocolumn[
\begin{@twocolumnfalse}
\maketitle
\begin{abstract}
Substrate strain offers a route to control deformation in supported metallic nanoparticles, provided that interfacial coupling transmits the imposed loading into the particle. We assess this feasibility using five static Cu/Si(111) configurations spanning 0--2\% biaxial tensile substrate strain. A 211,288-atom Cu particle is supported by a nominally $50\times50\times20$ nm Si slab with its bottom 1 nm fixed, using the Jelinek modified embedded-atom potential. Relative to the relaxed supported 0\% state, Cu-only local fits reveal heterogeneous, anisotropic deformation: particle-wide mean hydrostatic strain is negative at every nonzero loading, from $-3.12\times10^{-6}$ to $-8.84\times10^{-6}$, while the upper-particle mean varies nonmonotonically. Simulated Bragg CDI of Cu(111), obtained by inversion of known complex atomic scattering amplitudes, maps loading-dependent projected displacement and strain. The combined atomistic and Bragg CDI representations provide strong preliminary support for the feasibility of substrate-mediated strain transfer in this idealized system and identify spatial signatures for experimental investigation. Experiments remain indispensable to confirm transfer in real Cu/Si(111) nanoparticles and determine its magnitude.
\end{abstract}

\noindent\textbf{Keywords:} strain transfer; copper nanoparticle; Si(111); biaxial strain; hydrostatic strain; energy minimization; simulated Bragg CDI; PyNX; MEAM.
\vspace{0.8em}
\end{@twocolumnfalse}
]

\section{Introduction}
\label{sec:intro}
Straining a support offers a route to control deformation in an attached metallic nanoparticle. Interfacial coupling competes with free-surface relaxation, so a biaxially stretched substrate can generate different lateral and normal responses within the particle. Atomistic simulations of Co particles on expanded Cu substrates demonstrated spatially heterogeneous transfer, sensitive to temperature and interface structure~\cite{zhou2009}. For Cu on Si, studies of cluster deposition and film growth established the importance of epitaxy, intermixing, and preparation conditions~\cite{gong2016,zhu2023}. How prescribed substrate strain changes the deformation of an individual supported Cu particle remains the question addressed here.

The interface and surface termination are central to this response. Surface stress depends on the strain derivative of surface free energy~\cite{shuttleworth1950}, while Si(111) reconstruction~\cite{takayanagi1985} and its strain dependence~\cite{zhachuk2013} affect the atomic contact. We isolate substrate loading in a model with an initially ideal, oxide-free, unpassivated Si(111) surface, consistent particle preparation, and a fixed anchoring depth.

Bragg coherent diffraction imaging (BCDI) links nanocrystal deformation to a complex-object phase encoding the displacement projected along a reciprocal-lattice vector~\cite{pfeifer2006}. Its directional derivative gives the corresponding normal strain~\cite{gorobtsov2022}. Atlan et al.\ combined electrochemical BCDI of Pt with atomistic and density-functional modelling to resolve heterogeneous strain and propagation from surface to interior~\cite{atlan2023}. Chatelier et al.\ combined multireflection BCDI with atomistic modelling to study strained Ni$_3$Fe/Ni-rich core--shell formation~\cite{chatelier2024}. These studies provide a precedent for calculating Bragg scattering from atomic positions and analysing the resulting displacement and strain fields in a diffraction representation.

The present study connects three stages of this proposed control route: prescribed biaxial deformation of Si(111), the resulting static response of a supported Cu particle, and its representation by simulated Bragg CDI. The supports span 0\%, 0.5\%, 1\%, 1.5\%, and 2\% tensile strain. Hydrostatic distributions and directional tensor components characterize the Cu response relative to the relaxed supported 0\% state, while simulated Bragg CDI of the same coordinates identifies projected-displacement signatures.

The objective is a preliminary assessment of strain-transfer feasibility under a defined atomistic preparation protocol. Experimental confirmation is indispensable for establishing transfer in real Cu/Si(111) systems. The calculations provide a microscopic basis for that investigation by linking substrate loading to heterogeneous particle deformation and to candidate signatures in Bragg CDI. The analysis distinguishes hydrostatic and projected strain and states the numerical and model limits of this interpretation.

\section{Computational methodology}
\label{sec:methods}
\subsection{Geometry, initial state, and loading}
\begin{figure*}[!t]
\centering
\paperfigure{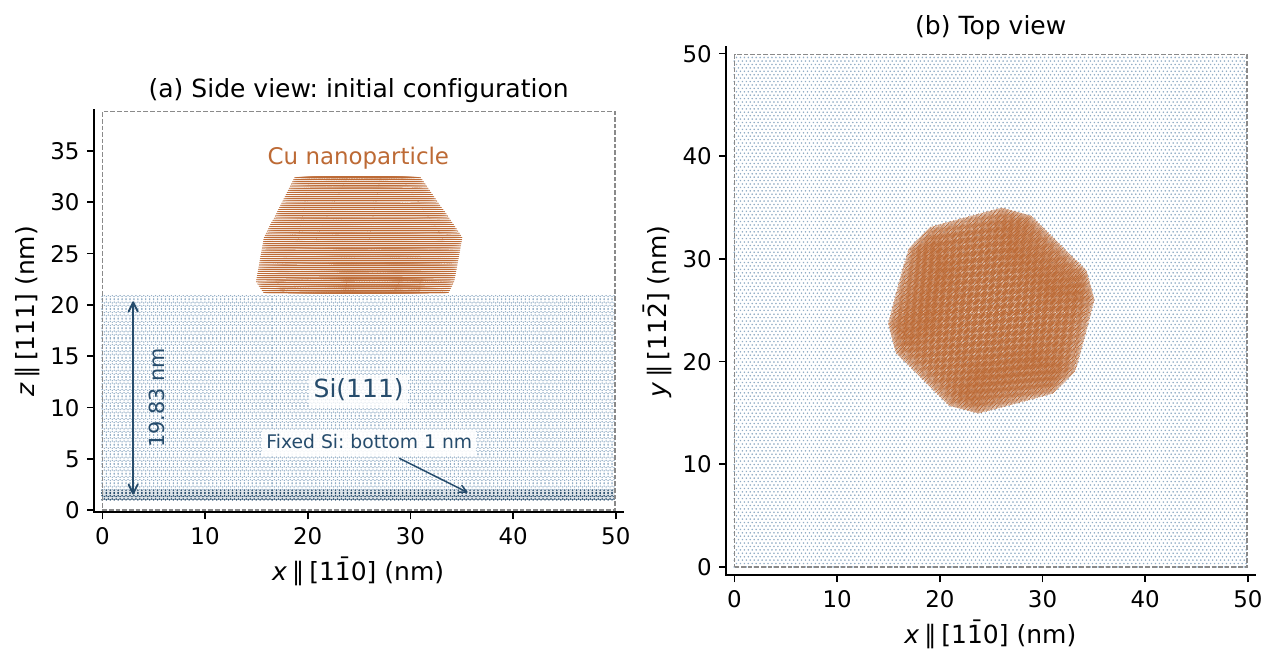}
\caption{Initial Cu nanoparticle on Si(111) at 0\% substrate strain. (a) Side view: Si is shown in a central 0.3 nm-thick slice and Cu in projection. The bottom 1 nm is fixed and the outermost Si-plane separation is 19.83 nm. (b) Top view in the $49.924\times49.887$ nm periodic lateral cell. Cu is orange, mobile Si light blue, and fixed Si dark blue. Dashed lines mark the simulation cell.}
\label{fig:geometry}
\end{figure*}

The model contains 211,288 Cu atoms and 2,496,000 Si atoms (Figure~\ref{fig:geometry}). The Cu particle has initial coordinate spans of $19.914\times19.914\times11.278$ nm and a construction lattice parameter of 3.615 \angstrom. Its atom count, crystalline geometry, orientation, and initial internal displacements are preserved across substrate strains. The diamond-cubic Si slab uses a reference lattice parameter of 5.431 \angstrom, with $x\parallel[1\bar10]$, $y\parallel[11\bar2]$, and $z\parallel[111]$. Its 64 (111) bilayers span 19.833 nm between outermost planes, corresponding to a 20.068 nm bulk-repeat thickness.

The reference lateral cell is $49.92386\times49.88692$ nm nad its height is 38.80707 nm. Boundaries are periodic in $x,y$ and nonperiodic in $z$. Si coordinates and lateral cell lengths are initially scaled by
\begin{equation}
\begin{aligned}
\mathbf F_{\mathrm{Si}}^{\mathrm{init}}&=\operatorname{diag}(1+\epsilon,1+\epsilon,1),\\
\epsilon&\in\{0,0.005,0.01,0.015,0.02\}.
\end{aligned}
\label{eq:loading}
\end{equation}
Cu is rigidly recentered, without rescaling, above Si with a 2.5 \angstrom\ initial gap. Cell dimensions remain fixed during relaxation. The bottom 1 nm of Si comprises 136,500 atoms in seven planes. Their coordinates remain fixed while their interactions with mobile atoms remain active. All other atoms are mobile. The supports are therefore prestrained before attachment, with normal Si relaxation allowed. Cell dimensions and construction details are listed in Supplementary Section S1 and Table S1.

\subsection{Interatomic potential and static relaxation}
LAMMPS~\cite{thompson2022} calculations use the modified embedded-atom method~\cite{baskes1992} with the Al--Si--Mg--Cu--Fe parameterization~\cite{jelinek2012} distributed by NIST~\cite{nist}. Conjugate-gradient relaxation holds the cell fixed, disables the energy-change criterion, and uses \texttt{min\_modify dmax 0.1 norm max}. All five states request $10^{-6}$ eV/\angstrom\ maximum per-atom force norm, with limits of 10,000 iterations and 100,000 force evaluations.

All calculations terminate with \texttt{linesearch alpha is zero}. Attained maximum mobile-atom force norms range from $1.07\times10^{-4}$ to $2.29\times10^{-4}$ eV/\angstrom\ (Supplementary Table S2), above the requested tolerance. The dataset comprises approximately relaxed, constrained athermal configurations. The implications for strain-transfer feasibility are discussed in Section~\ref{sec:experimentalscope}.

Cu is initially identical across loadings, whereas each Si support is prestrained before the combined configuration is relaxed. This preparation compares separately relaxed contacts and does not trace a quasistatic loading path of an already attached particle. The study addresses the static response of this preparation protocol.

\subsection{Reference-relative hydrostatic deformation}
\label{sec:strainmethod}
Full-precision minimized coordinates at 0\% define the reference. Atom IDs establish material correspondence, with no affine cell remapping. For reference and current neighbor vectors $\mathbf R_{ij}$ and $\mathbf r_{ij}$, the local least-squares deformation gradient~\cite{shimizu2007} and Green--Lagrange tensor are
\begin{equation}
\begin{aligned}
\mathbf F_i&=\underset{\mathbf F}{\operatorname{argmin}}\sum_{j\in\mathcal N_i^0}\|\mathbf r_{ij}-\mathbf F\mathbf R_{ij}\|^2,\\
\mathbf E_i&=\tfrac12(\mathbf F_i^{\mathrm T}\mathbf F_i-\mathbf I).
\end{aligned}
\label{eq:localfit}
\end{equation}
The primary reference neighborhood contains Cu atoms within 5 \angstrom. It therefore measures deformation within the metal rather than fitting a common affine transformation to metal and substrate atoms.

The hydrostatic and deviatoric parts are separated by
\begin{equation}
\begin{aligned}
h_i&=\tfrac13\operatorname{tr}(\mathbf E_i),\\
\mathbf E_i&=h_i\mathbf I+\mathbf E_i^{\mathrm{dev}},\quad
\operatorname{tr}(\mathbf E_i^{\mathrm{dev}})=0.
\end{aligned}
\label{eq:hydrostatic}
\end{equation}
Here $h_i$ is the three-dimensional scalar named \texttt{Volumetric Strain} in OVITO~\cite{ovitoatomic}. Its sign characterizes the isotropic part of the fitted tensor, without specifying the signs of individual components. For the lateral average $E_{\parallel,i}=(E_{xx,i}+E_{yy,i})/2$, the exact identity
\begin{equation}
h_i=\tfrac13(2E_{\parallel,i}+E_{zz,i})
\label{eq:balance}
\end{equation}
links the hydrostatic response to lateral and normal deformation. The identity also holds for means taken over the same material cohort with the same weights.

The local affine volume ratio is $J_i=\det\mathbf F_i$, and $J_i-1\simeq3h_i$ only in the infinitesimal limit. Neither $h_i$ nor an average of $J_i$ is a tessellated measurement of total particle volume. These kinematic quantities do not determine mechanical stress~\cite{lammpsstress} or separate elastic and plastic deformation. The reference is a relaxed supported particle, so negative $h_i$ denotes a change relative to that state, rather than a negative absolute stress relative to free Cu.

Statistics are evaluated for all Cu and for the 166,908 Cu atoms whose relaxed reference height is at least 2 nm above the lowest Cu atom. Fixed material cohorts preserve the same IDs across states. Reported quantities are the mean, median, positive fraction $f_+=N_{\mathcal R}^{-1}\sum_{i\in\mathcal R}\mathbf 1(h_i>0)$, distributions, and 1 nm height profiles. A global affine fit to centered Cu coordinates provides a particle-wide comparison. Mixed Cu/Si-neighbor fits and non-affine residuals probe interfacial sensitivity. Their definitions and numerical checks appear in Supplementary Section S2.

\subsection{Simulated Bragg CDI}
\label{sec:bcdimethod}
Only Cu atoms contribute to the coherent sum, evaluated with \texttt{pynx.scattering} (PyNX 2023)~\cite{pynx2020,pynxscattering}. For angular reciprocal coordinates $\mathbf Q$,
\begin{equation}
\begin{aligned}
A_\epsilon(\mathbf Q)&=\sum_{j=1}^{N_{\mathrm{Cu}}}w_j(\mathbf Q)e^{-i\mathbf Q\cdot\mathbf r_j(\epsilon)},\\
I_\epsilon(\mathbf Q)&=|A_\epsilon(\mathbf Q)|^2.
\end{aligned}
\label{eq:scatter}
\end{equation}
Here $w_j$ is an atomic scattering weight. PyNX coordinates in cycles per unit length satisfy $\mathbf Q=2\pi\mathbf s$. Phase conversion uses the angular convention. Each Cu point cloud is centered at its coordinate mean.

A $128^3$ Bragg volume is sampled using the nominal Cu lattice parameter 3.615 \angstrom. A centered forward Fourier transform of the conjugated complex amplitude recovers a real-space object. Reciprocal-space phase is therefore known in this simulation. The normalized-amplitude support threshold is 0.3, and the final crop retains $128^3$ voxels. Amplitude and phase receive the same center-of-mass shift, followed by nonperiodic phase unwrapping, linear-ramp removal on a 0.64 support, support-mean phase subtraction, and Fourier-domain Blackman filtering. Fourier signs, supports, and filtering affect quantitative interpretation~\cite{carnis2019} and are therefore paramount to the analysis. Complete processing and reciprocal-centroid records are given in Supplementary Section S4.

For the processed object phase $\phi_B$ and analysis vector $\mathbf Q_{Bragg}$,
\begin{equation}
u_{111}=\phi_B/|\mathbf Q_{Bragg}|,\qquad
\varepsilon_{111}=\widehat{\mathbf Q}_{Bragg}\cdot\nabla u_{111}.
\label{eq:projected}
\end{equation}
The conjugate-amplitude convention determines the displacement sign. In the small-deformation limit, the normal strain is $\widehat{\mathbf Q}_{Bragg}^{\mathrm T}\boldsymbol\varepsilon\widehat{\mathbf Q}_{Bragg}$, with $\boldsymbol\varepsilon$ the infinitesimal tensor. A single reflection resolves a projection, whereas $h$ is a tensor trace.

Differential displacement uses the 0\% wavevector norm for every loading,
\begin{equation}
\Delta\varepsilon_{\epsilon,0}=\widehat{\mathbf Q}_{d,\epsilon}\cdot\nabla\!\left[\frac{\phi_B^{(\epsilon)}-\phi_B^{(0)}}{|\mathbf Q_{Bragg,0}|}\right].
\label{eq:differential}
\end{equation}
The differentiation vectors, final-state voxel-size variables, and reused 2\% plotting support are detailed in Supplementary Section S4.

Removing a ramp $\boldsymbol\beta\cdot\mathbf r$ changes projected strain by the constant $-\widehat{\mathbf Q}_{Bragg}\cdot\boldsymbol\beta/|\mathbf Q_{Bragg}|$. Subtracting a constant phase only changes the displacement origin. Processed maps describe spatial variations with a fitted uniform offset removed.

\section{Results}
\label{sec:results}
\subsection{Imposed loading and relaxed substrate response}
All five systems retain their atom counts and fixed cell dimensions. At prescribed strains of 0.5\%, 1\%, 1.5\%, and 2\%, sampled interior Si gives mean $E_\parallel=0.50$\%, 1.01\%, 1.51\%, and 2.02\%, respectively. These values follow the Green--Lagrange conversion $[(1+\epsilon)^2-1]/2$. Normal means are $-0.235$\%, $-0.473$\%, $-0.714$\%, and $-0.959$\%. The slab retains lateral extension while contracting normally. Sampling and minimization records appear in Supplementary Section S1.

\subsection{Spatial redistribution of deformation in Cu}
\label{sec:hydroresponse}
\begin{figure*}[!t]
\centering
\paperfigure{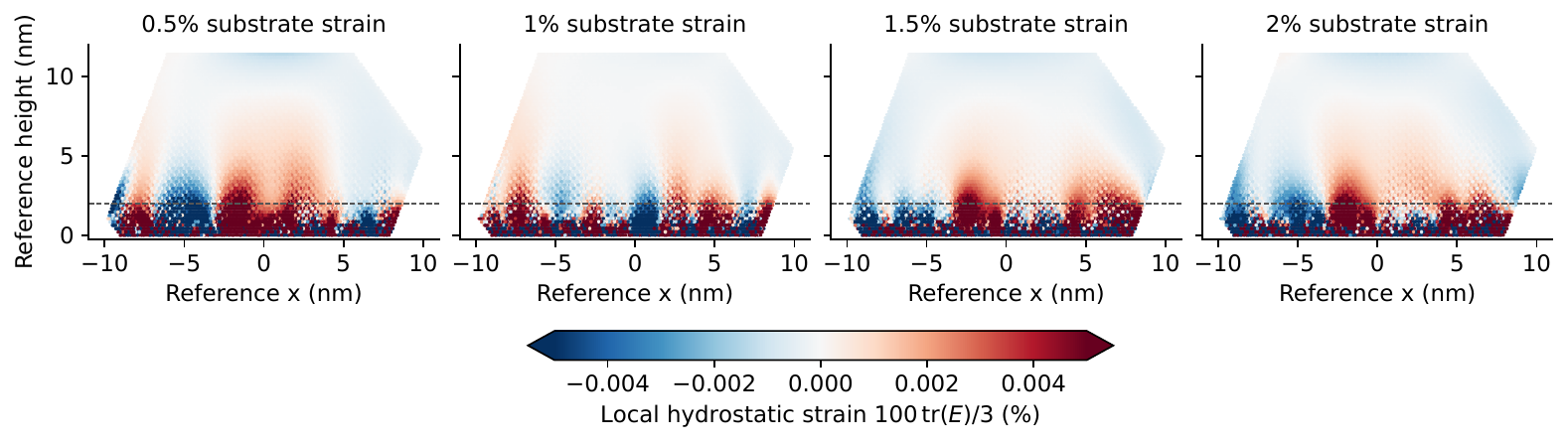}
\caption{Reference-relative Cu hydrostatic strain at 0.5\%, 1\%, 1.5\%, and 2\% substrate strain, from left to right. A central 1 nm-thick reference slice is shown in reference coordinates, relative to the relaxed 0\% state. Cu-only neighbors within 5 \angstrom\ define the tensor. The common scale saturates outside $\pm0.005$\%. The dashed line marks the 2 nm upper-cohort boundary. The complete 20 nm Si slab is omitted.}
\label{fig:maps}
\end{figure*}

Both hydrostatic signs coexist in every loaded configuration (Figure~\ref{fig:maps}). Cu-only particle-wide means are negative and become successively more negative across the sampled loadings: $-3.12$, $-5.36$, $-6.04$, and $-8.84$, in units of $10^{-6}$ dimensionless $h$ (Table~\ref{tab:strain}). Medians remain negative, whereas positive fractions vary nonmonotonically between 37.89\% and 44.51\%. These increments characterize changes in the supported Cu particle relative to its relaxed 0\% configuration.

\begin{table}[!t]
\centering\small
\setlength{\tabcolsep}{3pt}
\caption{Cu-only hydrostatic statistics at 5 \angstrom\ cutoff. $h$ is dimensionless. The upper cohort contains the same 166,908 reference atoms at height $\geq2$ nm.}
\label{tab:strain}
\begin{tabular}{@{}lrrrr@{}}
\toprule
Region & $\epsilon$ (\%) & $10^6\langle h\rangle$ & $10^6\mathrm{med}(h)$ & $f_+$ (\%)\\
\midrule
All Cu & 0.5 & $-3.123$ & $-1.480$ & 41.30\\
All Cu & 1 & $-5.365$ & $-0.517$ & 44.51\\
All Cu & 1.5 & $-6.040$ & $-2.111$ & 37.89\\
All Cu & 2 & $-8.841$ & $-1.573$ & 41.19\\
Above 2 nm & 0.5 & $-0.379$ & $-1.468$ & 39.28\\
Above 2 nm & 1 & $-0.588$ & $-0.468$ & 43.87\\
Above 2 nm & 1.5 & $-1.192$ & $-2.231$ & 33.82\\
Above 2 nm & 2 & $-0.701$ & $-1.591$ & 38.87\\
\bottomrule
\end{tabular}
\end{table}

\begin{figure*}[!t]
\centering
\paperfigure{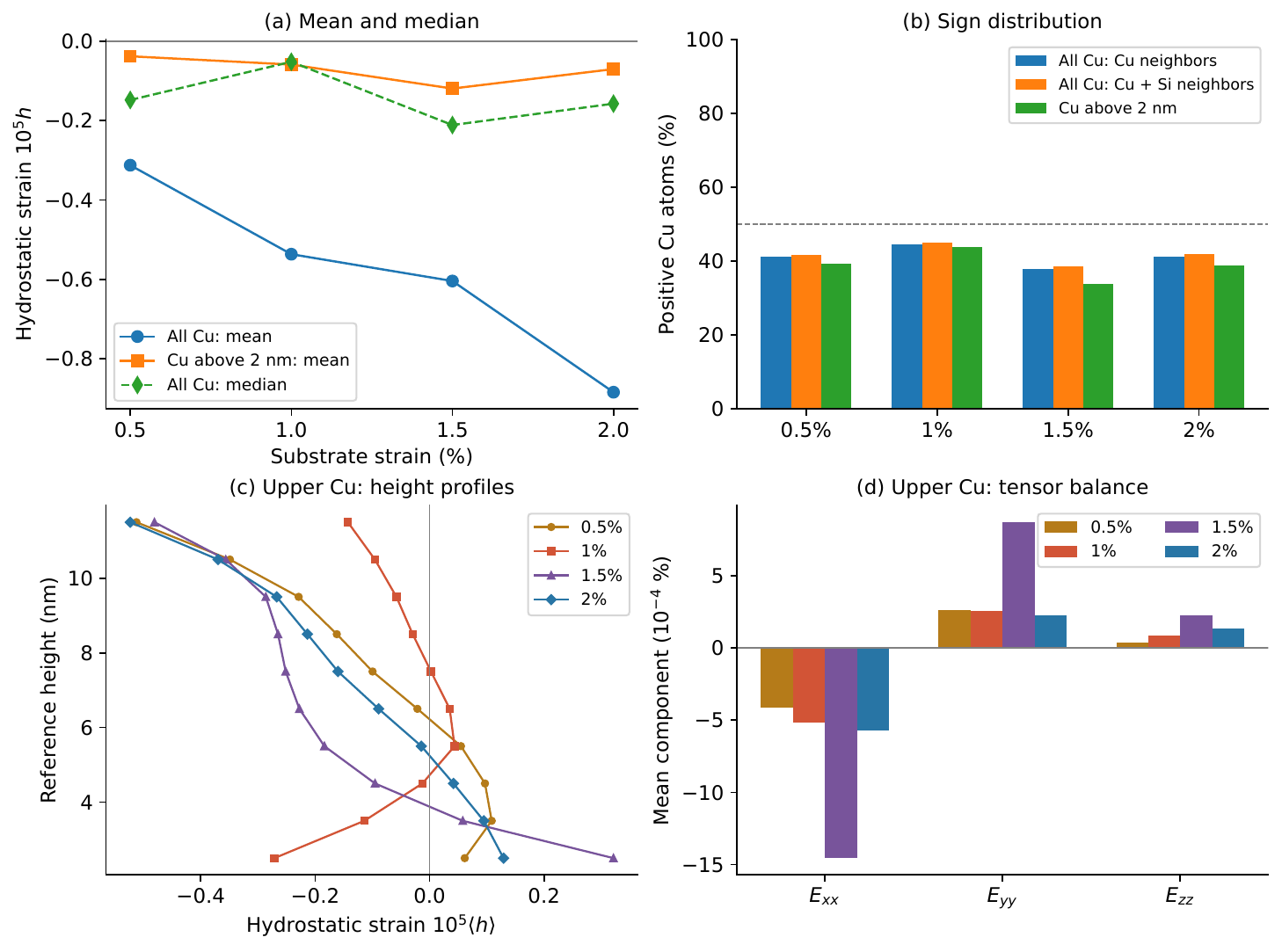}
\caption{Five-state static response. (a) Loaded-state mean and median $h$. (b) Positive fractions for Cu-only and mixed-species neighborhoods. (c) Cu-only mean $h$ in 1 nm reference-height bins above 2 nm. (d) Upper-Cu directional means. Lines connect separately relaxed configurations and do not define a quasistatic loading path or a critical strain. Spatial atom counts are not independent replicas.}
\label{fig:statistics}
\end{figure*}

The upper-Cu mean is smaller and nonmonotonic, with its most negative sampled value at 1.5\%: $-1.192\times10^{-6}$. Height profiles contain regions of both signs and distinguish the four loadings (Figure~\ref{fig:statistics}). The different loading dependence of particle-wide and upper-Cu means shows that deformation is redistributed spatially. The response cannot be represented by a uniform particle-wide contraction.

Global affine fits also give negative hydrostatic increments, but a different loading dependence: $-3.26$, $-4.12$, $-0.851$, and $-4.02$, in units of $10^{-6}$. A global fit weights spatial displacements differently from an average of local tensors. Its behavior confirms that a single particle-wide scalar does not characterize the full heterogeneous field.

\subsection{Anisotropic tensor balance and distributions}
All four upper-Cu cohorts have negative mean $E_{xx}$ and positive mean $E_{yy}$ and $E_{zz}$ (Table~\ref{tab:balance}). The hydrostatic mean is negative because contraction along $x$ outweighs expansion along $y$ and $z$. At 1.5\%, all three mean component magnitudes are largest among the sampled loadings: $E_{xx}=-14.55\times10^{-6}$, $E_{yy}=+8.71\times10^{-6}$, and $E_{zz}=+2.27\times10^{-6}$. The smaller component magnitudes at 2\% yield a less negative upper-Cu mean. The nonmonotonic hydrostatic response follows this balance of directional components.

\begin{table}[!t]
\centering\small
\setlength{\tabcolsep}{3pt}
\caption{Upper-Cu mean components in units of $10^{-6}$ dimensionless strain. Identical cohort weights satisfy $3\langle h\rangle=\langle E_{xx}+E_{yy}+E_{zz}\rangle$.}
\label{tab:balance}
\begin{tabular}{@{}rrrrr@{}}
\toprule
$\epsilon$ (\%) & $\langle E_{xx}\rangle$ & $\langle E_{yy}\rangle$ & $\langle E_{zz}\rangle$ & $\langle h\rangle$\\
\midrule
0.5 & $-4.160$ & $+2.644$ & $+0.380$ & $-0.379$\\
1 & $-5.167$ & $+2.535$ & $+0.868$ & $-0.588$\\
1.5 & $-14.553$ & $+8.708$ & $+2.271$ & $-1.192$\\
2 & $-5.716$ & $+2.265$ & $+1.346$ & $-0.701$\\
\bottomrule
\end{tabular}
\end{table}

\begin{figure*}[!t]
\centering
\paperfigure{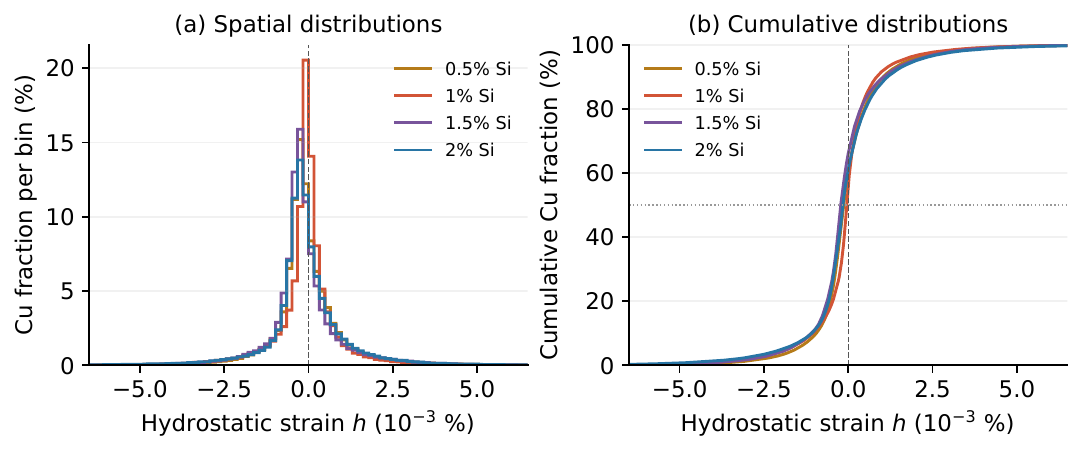}
\caption{Upper-Cu hydrostatic distributions for all four nonzero loadings. (a) Histogram bin counts divided by the full cohort. (b) Cumulative fractions. One horizontal unit is $10^{-3}$\%, or $10^{-5}$ dimensionless $h$. The common display window is $\pm0.007$\%, out-of-window fractions are 0.19\%, 0.29\%, 0.26\%, 0.57\% at 0.5\%, 1\%, 1.5\%, and 2\%, respectively. Full-cohort statistics are used in Table~\ref{tab:strain}.}
\label{fig:distributions}
\end{figure*}

The hydrostatic distributions overlap substantially (Figure~\ref{fig:distributions}). Upper-Cu spatial standard deviations range from 0.0011\% to 0.0014\%, much larger than the corresponding mean offsets. The 5th--95th percentile intervals are $[-0.0016,+0.0017]$\% at 0.5\% and $[-0.0018,+0.0019]$\% at 1.5\%, the endpoint 2\% interval is $[-0.002,+0.002]$\%. Mean, median, positive fraction, and width therefore describe different aspects of the response. Percentiles characterize spatial heterogeneity, not uncertainty in a replicated mean.

\subsection{Interfacial contribution and neighborhood dependence}
The contact region contributes disproportionately to particle-wide statistics. Mixed Cu/Si-neighbor fits give positive particle-wide means, while Cu-only means remain negative (Supplementary Table S3). For the 4,051-atom reference contact cohort, mixed-neighbor mean $h$ increases from 0.0498\% at 0.5\% loading to 0.55\% at 2\%. The corresponding Cu-only means are $-0.005$\% and $-0.016$\%. Mixed fits combine deformation within Cu with relative Cu--Si motion, and their non-affine residuals increase with loading. Above 2 nm, both neighborhood definitions yield identical tensors because Si lies outside the fitting radius. This separates deformation within the upper metal from affine-fit sensitivity at the junction. Full neighborhood and residual statistics are given in Supplementary Section S2.

\subsection{Signatures in simulated Bragg CDI}
\label{sec:bcdiresults}
\begin{figure*}[!t]
\centering
\paperfigure{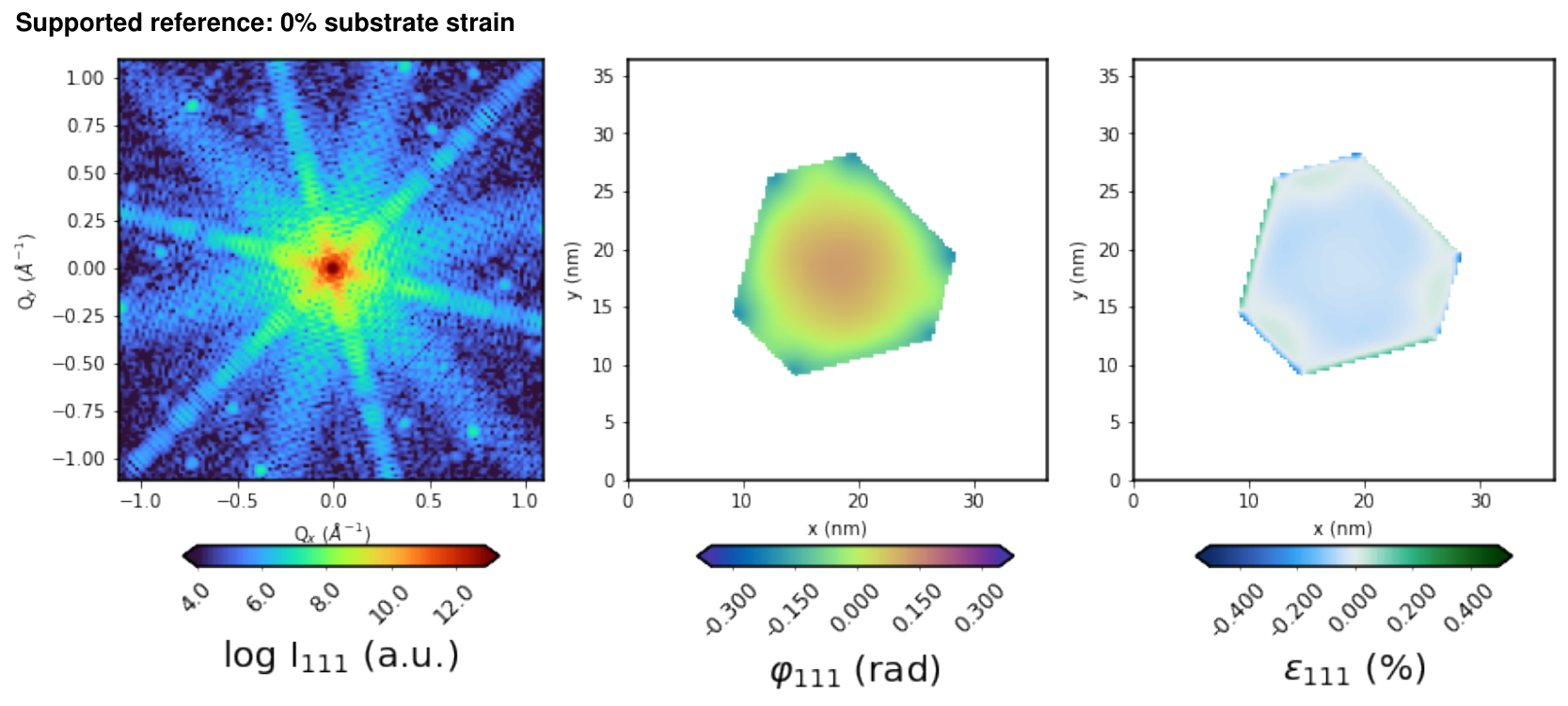}
\caption{Simulated Bragg CDI of Cu(111) for the supported 0\% reference: a reciprocal-space intensity section, complex-object phase section, and projected-strain section. Units are logarithmic arbitrary intensity, radians, and percent. This baseline is used to form differential fields. Full orthogonal sections for all five states appear in Supplementary Figures S1--S3.}
\label{fig:bcdi_overview}
\end{figure*}

The five Cu(111) Bragg volumes share a compact maximum and anisotropic fringes. Full phase fields and particle outlines are similar. Projected-strain sections retain a strong positive base-region band even at 0\% (Figure~\ref{fig:bcdi_overview}). Differential fields are therefore needed to isolate loading-dependent changes. Supplementary Figures S1--S3 document the complete five-state baseline comparison.

\begin{figure*}[p]
\centering
\paperfigure{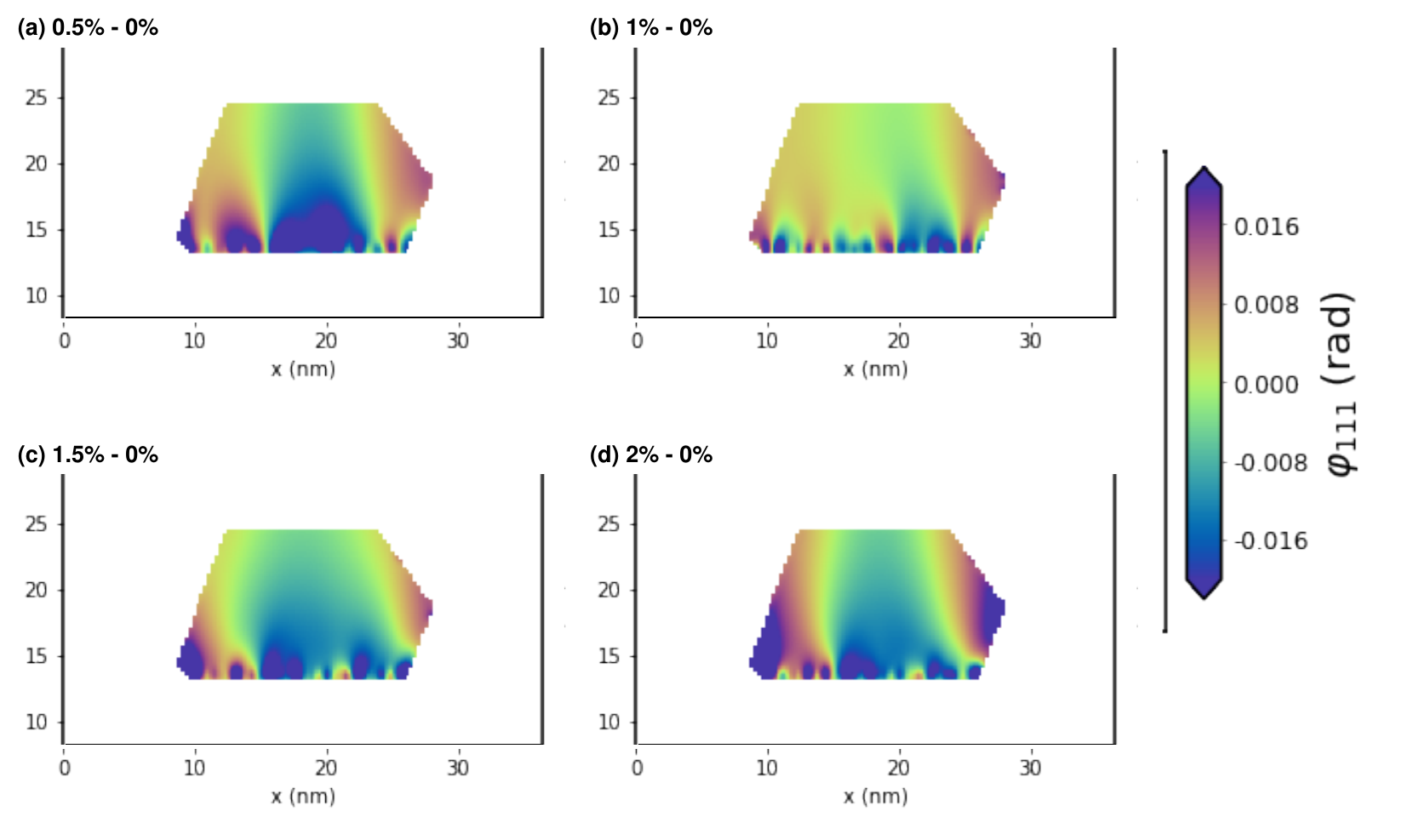}
\caption{Differential phase from simulated Bragg CDI of Cu(111), in radians: (a) 0.5\%--0\%, (b) 1\%--0\%, (c) 1.5\%--0\%, and (d) 2\%--0\%. Zoomed $xz$ sections retain the original coordinates and common color scale. Full $xy$, $xz$, and $yz$ sections are in Supplementary Figure S4.}
\label{fig:bcdi_dphi}
\vspace{5pt}
\paperfigure{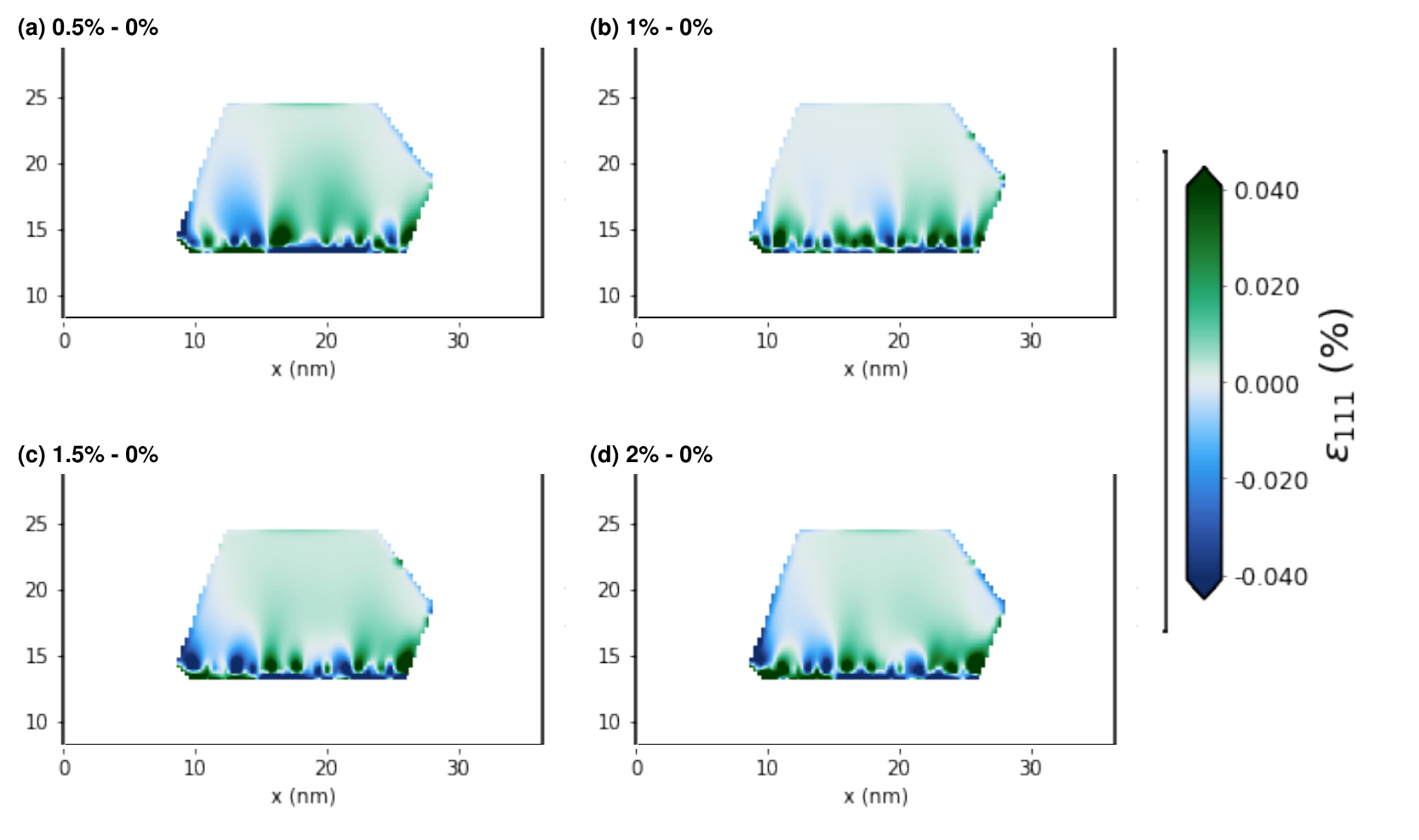}
\caption{Differential projected strain from simulated Bragg CDI, in percent, with the same state order and $xz$ sections. A fitted uniform phase-ramp contribution has been removed. This directional field differs from the atomistic hydrostatic scalar. Full orthogonal sections are in Supplementary Figure S5.}
\label{fig:bcdi_deps}
\end{figure*}

Differential phases vary spatially rather than following a uniform sign sequence (Figure~\ref{fig:bcdi_dphi}). At 0.5\%, a negative central region is pronounced near the lower part of the $xz$ section. The 1\% field is more laterally asymmetric, with mixed interior signs. At 1.5\% and 2\%, extended negative central regions coexist with positive lateral contrast and alternating features near the base. Full orthogonal sections show the corresponding three-dimensional heterogeneity (Supplementary Figure S4).

Differential projected strain contains both signs at every loading (Figure~\ref{fig:bcdi_deps}). Alternating base-region contrasts accompany weaker interior patches, whose locations and amplitudes change between states. Phase sign does not fix derivative sign, and neither field directly gives the hydrostatic scalar. The common differential display ranges are $\pm0.02$ rad and $\pm0.041$\% for phase and projected strain. These common ranges allow visual comparison of the loading-dependent spatial contrast.

The differential maps connect substrate loading to changes in the spatial distribution of projected Cu displacement and strain. Their relationship to the atomistic tensor, and the conditions for experimental confirmation, are discussed below.

\clearpage
\clearpage

\section{Discussion}
\label{sec:discussion}
\subsection{Feasibility of substrate-mediated strain transfer}
The model links prescribed Si loading to a deformation response in Cu. The substrate retains lateral extension and contracts normally, while the Cu particle develops loading-dependent tensor components and spatially heterogeneous hydrostatic fields. Cu is initially identical across loadings and is not directly rescaled. Its response arises through interactions with differently strained supports and their relaxed contacts. Deformation extends into the upper particle, beyond the immediate junction.

The central outcome is an anisotropic response to substrate loading. Contraction along one lateral direction coexists with expansion along the other lateral and normal directions. Particle-wide hydrostatic means become more negative across the sampled loadings, whereas upper-Cu means are nonmonotonic. The 1.5\% configuration has the most negative sampled upper-Cu mean, but does not define a critical strain or a uniform transition. These findings support a picture in which the interface and free surfaces redistribute the imposed deformation.

Combined with the loading-dependent simulated Bragg CDI fields, this model response provides strong preliminary support for the feasibility of substrate-mediated strain transfer in supported Cu nanoparticles. It gives a concrete microscopic basis for experimental investigation. Registry changes, slip, surface relaxation, and defects are possible contributors to the anisotropy. Structural analysis such as polyhedral template matching~\cite{larsen2016} would help distinguish them.

\subsection{Connecting hydrostatic deformation to simulated Bragg CDI}
Hydrostatic strain and a single-reflection strain describe complementary aspects of the same deformation. In the infinitesimal limit, write $\boldsymbol\varepsilon=h_{\mathrm{lin}}\mathbf I+\boldsymbol\varepsilon^{\mathrm{dev}}$. The normal strain projected along the Bragg vector is
\begin{equation}
\varepsilon_{111}=h_{\mathrm{lin}}+\widehat{\mathbf Q}_B^{\mathrm T}\boldsymbol\varepsilon^{\mathrm{dev}}\widehat{\mathbf Q}_B.
\label{eq:projectionbalance}
\end{equation}
The deviatoric contribution can outweigh the hydrostatic term. Consequently, a negative hydrostatic mean can coexist with positive projected strain, and heterogeneous displacement can remain appreciable when the scalar mean is small. Using the finite-strain scalar in this relation is an approximation appropriate to weak interior deformations, with additional care required at rearranged contact sites.

Simulated Bragg CDI translates the atomistic response into projected-displacement and strain contrast. Differential maps isolate loading-dependent spatial changes from the supported 0\% baseline and identify candidate regions for experimental comparison. Both analyses derive from the same coordinates and provide complementary representations within the model. A quantitative comparison requires matched physical grids, spatial resolution, supports, and phase references. Linear-ramp removal changes a uniform projected-strain offset. The processed maps therefore emphasize spatial variations. Supplementary Section S4 records these conventions.

\subsection{Experimental confirmation and scope}
\label{sec:experimentalscope}
Experiments are indispensable to confirm strain transfer in real Cu/Si(111) nanoparticles and determine its magnitude. A direct test would impose and independently measure the substrate strain while tracking the response of the same supported particle by Bragg CDI, referenced to an unloaded state. Multireflection measurements could resolve directional responses, building on combined atomistic and BCDI approaches~\cite{atlan2023,chatelier2024}. Experimental confirmation would establish whether the predicted response persists with the actual surface termination, contact structure, and achievable measurement sensitivity.

The present assessment remains preliminary. All minimizations stop by line-search termination above their common force target, so stability of the weak hydrostatic means and the nonmonotonic upper-Cu response can be challenged. The ideal oxide-free interface, anchoring depth, finite periodic cell, empirical potential, and separately prepared contacts define the model scope. The known-amplitude inversion does not establish recovery from experimental intensity alone, and the present images do not determine a transfer efficiency or experimental detection threshold. These limits constrain quantitative extrapolation. Within this scope, the model defines an experimentally testable link between substrate strain and particle deformation.

\section{Conclusions}
Five static Cu/Si(111) configurations spanning 0--2\% biaxial substrate strain show loading-dependent, anisotropic deformation in a supported Cu particle. Cu-only hydrostatic means are negative at all nonzero loadings, with coexisting positive and negative regions and a nonmonotonic upper-particle response. Simulated Bragg CDI of Cu(111) maps the associated redistribution of projected displacement and strain, connecting the atomistic response to candidate experimental signatures.

Together, these preliminary results provide strong support for the feasibility of substrate-mediated strain transfer in the idealized Cu/Si(111) system and motivate its experimental investigation. Experiments are indispensable to confirm transfer in real nanoparticles and establish its magnitude and reproducibility. Numerical convergence, realistic interface conditions, and quantitative treatment of projected fields will determine how far the present predictions extend to such measurements.

\section*{Supplementary materials}
Supplementary Figures S1--S5 show complete orthogonal fields. Supplementary Sections S1--S4 document geometry, static records, the potential, and processing conventions.

\section*{Acknowledgments}
The author acknowledges the CEA for providing access to high-performance computing (HPC) resources. The author thanks Dr. C. Atlan, Dr. E. Polack and Dr. E. Bellec for the fruitful discussions.

\end{document}


\maketitle

\noindent Supplementary Figures S1--S5 are provided in the separate simulated Bragg CDI figure supplement. The present document contains Tables S1--S3 and the static numerical records.

\section{Construction and minimization records}
The diamond-cubic slab contains 64 complete (111) bilayers, or 128 planes, using 130 and 75 repeats of an orthorhombic surface cell. The initial empty-space thicknesses below Si and above Cu are 1 nm and approximately 6.446 nm, respectively. The fixed 136,500-atom region is held fixed while its interactions with mobile atoms remain active. There are 2,359,500 mobile Si atoms and 211,288 Cu atoms. Only Si and the lateral cell are scaled at initialization. The Cu particle is translated without deformation. Cell dimensions remain fixed during minimization.

\begin{table}[htbp]
\centering
\caption{Prescribed substrate engineering strain and fixed lateral cell dimensions. All five cases contain 2,707,288 atoms and have $L_z=38.807$ nm.}
\label{tab:cells}
\begin{tabular}{rrr}\toprule
$\epsilon$ (\%) & $L_x$ (nm) & $L_y$ (nm)\\\midrule
0 & 49.924 & 49.887\\
0.5 & 50.173 & 50.136\\
1 & 50.423 & 50.386\\
1.5 & 50.673 & 50.635\\
2 & 50.922 & 50.885\\
\bottomrule\end{tabular}
\end{table}

Conjugate-gradient minimization uses the same $10^{-6}$ eV/\angstrom\ force target, disabled energy-change criterion, 10,000-iteration limit, and 100,000-evaluation limit in all states. \texttt{min\_modify dmax 0.1 norm max} selects the maximum per-atom norm. Every run stops with \texttt{linesearch alpha is zero}. None reaches the requested tolerance. Coordinate headers identify LAMMPS 29 October 2020 and agree with iteration counts.

\begin{table}[htbp]
\centering
\caption{Five-state static relaxation records. $F_{\max}$ is the maximum mobile-atom force norm in eV/\angstrom, distinct from the maximum component and global two-norm in the minimizer summary.}
\label{tab:min}
\begin{tabular}{rrrr}\toprule
$\epsilon$ (\%) & Iterations & Force evaluations & Attained $F_{\max}$\\\midrule
0 & 1619 & 3264 & $1.261\times10^{-4}$\\
0.5 & 2015 & 4056 & $1.416\times10^{-4}$\\
1 & 2195 & 4406 & $1.371\times10^{-4}$\\
1.5 & 2334 & 4693 & $1.070\times10^{-4}$\\
2 & 2374 & 4764 & $2.290\times10^{-4}$\\
\bottomrule\end{tabular}
\end{table}

Final potential energies, in loading order 0\%, 0.5\%, 1\%, 1.5\%, and 2\%, are $-12 258 107.479$, $-12 256 947.152$, $-12 252 364.133$, $-12 244 535.447$, and $-12 233 633.033$ eV. Their differences include Si strain energy and are not isolated adhesion energies. Interior Si is sampled at the same 16,607 reference sites with $5\leq z\leq15$ nm, at least 15 nm from the lateral center and away from periodic edges, using Si-only 5 \angstrom\ neighborhoods. Normal contraction describes the anchored slab rather than a fitted bulk Poisson ratio.

\section{Neighborhood sensitivity and numerical checks}
Configurations are matched by atom ID without affine cell remapping. Cu fitting neighborhoods do not cross periodic boundaries. All 211,288 Cu fits are sufficiently populated and nonsingular, for both neighbor definitions. Known affine deformation, rigid rotation, and reference-state tests yield tensor errors below $10^{-14}$. A global affine fit to centered Cu coordinates removes translation. These checks verify the estimator and do not bound minimization error.

\begin{table}[htbp]
\centering
\caption{Neighborhood sensitivity for all Cu central atoms at 5 \angstrom\ cutoff. The mixed-species mean is positive, while medians and majority signs remain negative.}
\label{tab:neighbors}
\begin{tabular}{lrrrr}\toprule
Neighbors & $\epsilon$ (\%) & $10^6\langle h\rangle$ & $10^6\operatorname{median}(h)$ & $f_+$ (\%)\\\midrule
Cu only & 0.5 & $-3.123$ & $-1.480$ & 41.30\\
Cu only & 1 & $-5.365$ & $-0.517$ & 44.51\\
Cu only & 1.5 & $-6.040$ & $-2.111$ & 37.89\\
Cu only & 2 & $-8.841$ & $-1.573$ & 41.19\\
Cu and Si & 0.5 & $+9.695$ & $-1.434$ & 41.63\\
Cu and Si & 1 & $+31.346$ & $-0.475$ & 45.04\\
Cu and Si & 1.5 & $+67.054$ & $-2.017$ & 38.59\\
Cu and Si & 2 & $+113.205$ & $-1.461$ & 42.02\\
\bottomrule\end{tabular}
\end{table}

The contact cohort contains 4,051 reference Cu atoms with nearest-Si distance at most 3.5 \angstrom. At 0.5\%, 1\%, 1.5\%, and 2\%, mixed-neighbor contact mean $h$ is 0.04977\%, 0.15839\%, 0.32156\%, and 0.55168\%. Cu-only means are $-0.00497$\%, $-0.00707$\%, $-0.01435$\%, and $-0.01627$\%. The unnormalized residual $D_{\min,i}^2=\sum_j\|\mathbf r_{ij}-\mathbf F_i\mathbf R_{ij}\|^2$ has mixed-neighbor contact medians 0.537, 2.04, 4.43, and 8.14 \angstrom$^2$, compared with Cu-only values 0.0438, 0.102, 0.157, and 0.219 \angstrom$^2$. Neighbor count affects this sum. Upper-Cu medians range from $3.21\times10^{-8}$ to $3.96\times10^{-8}$ \angstrom$^2$. These residuals describe affine-fit quality and do not uniquely identify a defect mechanism.

\section{Potential provenance and reproduction}
Calculations use \texttt{units metal}, a 5 \angstrom\ MEAM cutoff, a 2 \angstrom\ neighbor-list skin, and atomic masses of 63.546 u for Cu and 28.0855 u for Si. The potential-library order is \texttt{AlS SiS MgS CuS FeS}, with atom types 1 and 2 mapped to \texttt{CuS} and \texttt{SiS}. The interaction mapping is

\begin{verbatim}
pair_style meam
pair_coeff * * potentials/Jelinek_2012_meamf &
               AlS SiS MgS CuS FeS &
               potentials/Jelinek_2012_meam.alsimgcufe CuS SiS
\end{verbatim}

\section{Simulated Bragg CDI processing and comparison conventions}
The Cu-only calculation uses \texttt{pynx.scattering} (PyNX 2023)~\cite{pynx2020,pynxscattering}. All five coordinate point clouds match their relaxed Cu configurations bijectively after mean-coordinate centering, with maximum differences below $10^{-10}$ \angstrom\ despite reordered rows. The workflow specifies a $128^3$ grid, lattice parameter 3.615 \angstrom, and scattering-helper argument \texttt{nstep=100}. A nominal $[111]$ vector is rotated toward plotting $z$ before the call. Reciprocal spacing and coordinate transformation depend on the helper implementation.

Known complex \texttt{fhkl} is conjugated and inverted through a centered forward Fourier transform. The primary support threshold is 0.3 and the final crop $128^3$. Amplitude COM defines identical phase and amplitude shifts. Unwrapping uses seed 1 and nonperiodic boundaries. Ramp fitting uses a 0.64 support. Primary-support mean phase subtraction precedes Blackman filtering of the complex object. These steps specify simulated Bragg CDI from known amplitudes rather than intensity-only phase retrieval.

Intensity centroids, in \angstrom$^{-1}$, are
\begin{align*}
\mathbf Q_{c,0\%}&=(-0.00006,-0.00011,3.00490),\\
\mathbf Q_{c,0.5\%}&=(-0.00037,-0.00007,3.00490),\\
\mathbf Q_{c,1\%}&=(-0.00021,-0.00006,3.00489),\\
\mathbf Q_{c,1.5\%}&=(-0.00024,+0.00006,3.00491),\\
\mathbf Q_{c,2\%}&=(-0.00006,-0.00018,3.00491).
\end{align*}
Peak maxima are $(0,0,3.01046)$ in every state. These rounded outputs do not resolve relative lattice changes of order $10^{-6}$. The returned \texttt{q\_hkl} is used for normal strain, and \texttt{q\_com} is recorded for each system. Real-space COM translation is a separate operation.

Every differential displacement is divided by \texttt{q\_norm\_0}. The 0.5\% and 1\% derivatives use \texttt{q\_hkl\_0}, while the 1.5\% and 2\% derivatives use the final-state \texttt{q\_hkl} variable from 2\%. Voxel sizes and plotting support also reuse the 2\% variables. An identical nominal grid and printed peak position do not by themselves verify numerical equality of every reciprocal vector and physical voxel size. The support is not an explicit intersection of the two compared states. Full-state phase scales are automatic. Strain and differential scales are common.

Main Figures 6 and 7 display zoomed $xz$ sections through PDF viewports of the original PNGs. The complete particle in that section, original coordinate labels, and common color scale are retained. Unused empty plot space is omitted. No pixels are recolored and no strain values are inferred from images. Supplementary Figures S4 and S5 preserve all three complete sections for each differential state.

Quantitative comparison requires full reciprocal arrays, voxel sizes, numerical helpers, derivative and boundary conventions, exact PyNX patch release, and retained phase ramps. These fields are absent from the embedded PNGs. Shared material support, frame, resolution, normalization, and phase references must be established before comparing projected voxel fields. No simulated Bragg CDI averages, uncertainty estimates, or transfer coefficients are digitized from image colors.